\documentclass[submission, Phys]{SciPost}

\usepackage[english]{babel}
\usepackage{amsmath}
\usepackage{graphicx}
\usepackage{amsfonts}
\usepackage{amssymb}
\usepackage{graphicx}
\usepackage{mathrsfs}
\usepackage{layout}
\usepackage{hyperref}
\usepackage{mathtools}
\usepackage{stmaryrd}
\usepackage{url}
\usepackage{titlesec}
\usepackage{bbold}
\usepackage{caption}
\usepackage{subcaption}
\usepackage{lineno}

\definecolor{medgreen}{rgb}{0.2,0.6,0.2}

\usepackage{tikz}
\usetikzlibrary{patterns}
\usetikzlibrary{calc}
\usetikzlibrary{decorations.pathreplacing,decorations.markings}
\usetikzlibrary{arrows}

\newcommand{\tr}{\text{tr}}
\newcommand{\p}{\partial}
\newcommand{\vx}{\mathbf{x}}
\newcommand{\vxp}{\vx'}
\newcommand{\vq}{\mathbf{q}}

\newcommand{\vp}{\mathbf{p}}
\newcommand{\vv}{\mathbf{v}}

\newcommand{\vf}{\mathbf{f}}

\newcommand{\vnabla}{\mathbf{\nabla}}
\newcommand{\tj}{\tilde{j}}
\newcommand{\tih}{\tilde{h}}
\newcommand{\tvarphi}{\tilde{\varphi}}

\newcommand{\bla}{\big\langle}
\newcommand{\bra}{\big\rangle}

\newcommand{\nbR}{\mathbb{R}}
\newcommand{\grad}{\mathbf{\nabla}}
\begin{document}

\begin{center}{\Large \textbf{
The non-perturbative sides of the Kardar-Parisi-Zhang equation
}}\end{center}

\begin{center}
 L\'eonie Canet$^1$\\
 {\small $^1$ Univ. Grenoble Alpes, CNRS, LPMMC, 38000 Grenoble, France}
\end{center}

\vspace{5pt}

\begin{center}
\begin{minipage}{14cm}
{\small
The Kardar-Parisi-Zhang (KPZ) equation is a celebrated non-linear stochastic dynamical equation yielding non-equilibrium universal scaling. It exhibits notorious non-perturbative aspects. The KPZ fixed point is strong-coupling, all the more in $d>1$. Strikingly,  another, even  stronger-coupling fixed point of the KPZ equation, called inviscid Burgers fixed point, has been recently unveiled. These non-pertubative features can be theoretically accessed and studied in a controlled way in all dimensions using the functional renormalisation group. We propose an overview of the related results, which provide a  unified picture of the fixed-point structure and  associated  scaling regimes of the KPZ equation in $d=1$ and in higher dimensions.
}
\end{minipage}
\end{center}

%\vspace{10pt}
%\noindent\rule{\textwidth}{1pt}
%\tableofcontents\thispagestyle{plain}
%\noindent\rule{\textwidth}{1pt}
\vspace{5pt}

\section{Introduction}
\label{sec:intro}

The Kardar-Parisi-Zhang (KPZ) equation has become a fundamental model of statistical physics for self-organised criticality and non-equilibrium scaling phenomena. It models the stochastic dynamics of a growing interface, described by a singled-valued height field $h(t,\vx)$, as
\begin{equation}
\label{eq:KPZ}
    \partial_t h = \nu \nabla^2 h + \frac{\lambda}{2}(\nabla h)^2 + \eta\,,
\end{equation}
where $t$ is time, $\vx\in \nbR^d$ is the coordinate of the $d$-dimensional interface embedded in the space of dimension $d+1$ in which the growth takes place. The parameters are $\nu$ the surface tension,  $\lambda$  the strength of the non-linearity, and $D$ the amplitude of the  Gaussian noise $\eta$, which is of zero mean and of covariance
\begin{equation}
\label{eq:noise}
\bla \eta(t,\vx) \eta(t',\vxp)\bra = 2 D\delta(t-t')\delta^d(\vx-\vxp)\,.
\end{equation}
Note that time and field can be rescaled such that the KPZ equation has only one free parameter which is the effective non-linearity $g=\lambda^2 D/\nu^3$.
The KPZ dynamics~\eqref{eq:KPZ} generically drives the system to a critical state, characterised by universal power-law behaviours with anomalous exponents. Notably, the two-point correlation function endows the scaling form
\begin{equation}
\label{eq:correlation}
 C(t,\vx) \equiv \bla h(t,\vx)h(0,\mathbf{0})\bra_c \sim |\vx|^{2\chi} \bar{F}_{\rm KPZ}\big(|\vx|/t^{1/z}\big)\,
\end{equation}
where the subscript $c$ stands for connected, $\chi$ and $z$  are  the universal roughness and dynamical exponents.  The scaling function $\bar{F}_{\rm KPZ}$ is also universal, but depends on the type of initial conditions for the interface,  which define three main universality subclasses corresponding to flat, curved, or stationary initial configurations. While these subclasses share  the same values  for the critical exponents, they are each associated with a different universal form for the scaling function,  and also for the probability distribution of the rescaled height fluctuations~\cite{Corwin12,Takeuchi18}. We here consider the  stationary interface.

The KPZ equation exactly maps to the  Burgers equation for randomly stirred fluids~\cite{Burgers48,Forster77}
\begin{equation}\label{eq:Burgers}
    \partial_t \vv + \vv \cdot \vnabla \vv  = \nu \nabla^2 \vv + \vf\,,
\end{equation}
upon defining $\vv = -\lambda \grad h$ and $\vf = -\lambda\grad \eta$, where $\nu$  now embodies the kinematic viscosity of the fluid. The KPZ equation is also related to other celebrated  models of statistical physics, such as the equilibrium problem of directed polymers in random media~\cite{Kardar87}, or driven-dissipative transport and non-linear fluctuating hydrodynamics~\cite{Beijeren85,Janssen86,Spohn2014}.
 Beyond these connections, the KPZ universality class has turned out to be extremely large, and, remarkably, it keeps extending. Indeed, the KPZ universal behaviour was first evidenced in growth problems, such as combustion front propagation~\cite{Maunuksela97}, bacterial or cancer cells growth~\cite{Wakita97,Huergo2012}, and even  urban skylines~\cite{Najem2020}. The most accurate experiments have been realised in turbulent nematic liquid crystals where the finest KPZ properties could be measured~\cite{Takeuchi10,Takeuchi12}. We refer the reader to the nice existing reviews for a detailed account~\cite{Halpin-Healy95,Barabasi95,Krug97,Takeuchi18,Halpin-Healy15}.
 More surprisingly, the KPZ universality has also turned out to be relevant in quantum systems, where it has recently been observed both in integrable systems, such as Heisenberg quantum spin chains~\cite{Bloch2022,Tennant2022}, and in driven-dissipative systems, such as exciton-polariton condensates~\cite{Fontaine2022Nat}.
This has sparked a renewed  interest in this seminal equation.

 In dimension $d=1$, the interface always roughens, and exhibits a superdiffusive dynamics characterised by a critical dynamical exponent $z=3/2$, which  distinguishes it from the linear growth (described by~\eqref{eq:KPZ} with $\lambda=0$ which is the Edwards-Wilkinson (EW) equation~\cite{Edwards82}). The EW equation leads instead to a diffusive behaviour $z=2$, although sharing in $d=1$ the same roughness exponent $\chi=1/2$. Beyond the critical exponents, the statistical properties of the KPZ interface in $d=1$ are known to an incredibly  fine level, including the precise characterisation of the three universality subclasses, thanks to joined efforts from both the mathematical and theoretical physics communities.  Comprehensive reviews of these advances can be found in {\it eg} Refs~\cite{Corwin12,Takeuchi18}. Yet, the one-dimensional KPZ equation still holds its surprises, as we explain below.

 Let us first comment on dimensions $d>1$, where no exact results are available. While for $d\leq 2$, the interface always becomes rough, there exists in $d>2$ a phase transition, called roughening transition (RT) and corresponding to a critical non-linearity $\lambda_c$, which separates a smooth  EW phase for $\lambda <\lambda_c$ from a rough KPZ phase for $\lambda >\lambda_c$.  While the exponents of the linear EW phase can be simply determined as $\chi=(2-d)/2$ and $z=2$, the exponents of the KPZ phase are not known exactly, and have been mainly  estimated through numerical simulations.
 As emphasised previously, the rough interface is characterised by scale invariance and anomalous exponents. A choice method to understand critical behaviours, which has proved pivotal to deal with such situations in equilibrium, is the renormalisation group (RG), as pioneered by Wilson~\cite{Wilson74}. The idea of the RG
  is to perform the average over fluctuations not at once, but scale by scale, thereby  building scale-dependent effective theories. The evolution of these theories with the scale is given by a RG flow equation. A key feature of the RG procedure is that when the system is scale invariant, the RG flow reaches a fixed point, and the effective description converges to a fixed one. This is analogous to a dynamical system,  where the role of time is played by the RG scale, the physical phase space is replaced by an abstract theory space, and the fixed points of the dynamics, which describe stationary (time-independent) solutions, are superseded by  fixed points of the RG flow, which describe scale-invariant solutions.

 For the KPZ equation, the RG flow  was early derived at one-loop order~\cite{Forster77,Kardar86} and yields in $d=1$ the exact exponents $\chi=1/2$ and $z=3/2$. It was shown later that this flow could in fact be obtained to all orders in the perturbative expansion in $\lambda$, and resummed~\cite{Wiese98}. However, although it allows one to describe exactly the RT fixed point controlling the roughening  phase transition, it fails to find the KPZ fixed point in $d\geq 2$, and leads instead to a flow running away to infinity.  The KPZ rough phase is thus genuinely non-perturbative, since it cannot be accessed at any order of perturbation theory. To make progress, one needs to resort to a non-perturbative approach, and the functional RG (FRG) provides such a tool. This method has indeed allowed one to find the KPZ fixed point in all dimensions~\cite{Canet2010,Kloss2012}. Moreover, it is very  general and versatile, such that it can be used in other non-integrable cases, such as in the presence of microscopic correlations of the noise, spatial~\cite{Kloss2014a,Mathey2017} or temporal~\cite{Squizzato2019}, and in anisotropic settings~\cite{Kloss2014b}.

As already hinted at, the non-perturbative side of the KPZ equation surprisingly also arises in $d=1$. Indeed, recent numerical simulations have unveiled an unexpected scaling regime, characterised by a dynamical exponent $z=1$, which is absent from the exact results. This scaling has been observed while approaching the inviscid (zero-viscosity) limit of the stochastic Burgers equation~\cite{Brachet2022}, as well as  the equivalent  tensionless limit of the KPZ equation~\cite{Rodriguez2022}. Let us notice that,  intriguingly, the same scaling  has also been evidenced in a strongly interacting one-dimensional quantum bosonic  model (the Bose-Hubbard model), although the connection with KPZ dynamics is not clearly established in this case~\cite{Fujimoto2020} \footnote{The authors of Ref.~\cite{Fujimoto2020} report, at high filling, a collapse of their data onto a scaling form~\eqref{eq:correlation} consistent with the EW one. A crossover from the EW to the asymptotic KPZ scaling is known to occur when the non-linearity is small and the system size not large enough compared to the length-scale set by this nonlinearity~\cite{Nattermann92}. This suggests that, although a formal mapping to the KPZ equation is not available for the Bose-Hubbard model, it may correspond, at large filling, to the KPZ universality class with a small non-linearity. The authors then find, at half-filling, a change from the EW universal behaviour to a different one termed ``unconventional universality class'', which  features the dynamical exponent $z=1$ similar to the IB scaling, providing a further support for the  possible connection of this model with KPZ dynamics.}. This unexpected scaling regime suggests the existence of a new fixed point of the KPZ equation. This fixed point  has indeed been  uncovered using FRG, and has  been termed the inviscid Burgers (IB) fixed point~\cite{Fontaine2023InvBurgers,Gosteva2024}.
 The purpose of this paper is to give an overview of these non-perturbative aspects, and show how they can be addressed and characterised using FRG. It does not contain new results, but aims at providing a unified picture of the existing results, in a comprehensive  and non-technical way.

 The remainder of the paper is organised as follows. The bases of the FRG formalism  and its formulation for the KPZ equation are set in Sec.~\ref{sec:FRG}. We  show in Sec.~\ref{sec:KPZ} how the KPZ fixed point can be captured in all $d$ with the simplest approximation of the FRG, and then explain how quantitative results can be obtained within this formalism. In Sec.~\ref{sec:IB}, we focus on the inviscid limit and show that the IB fixed point can  also be identified within the simplest approximation of the FRG, yielding the complete phase diagram of the KPZ equation. We then provide a quantitative characterisation of this fixed point.

\section{Functional renormalisation group for the KPZ equation}
\label{sec:FRG}

The starting point of the FRG is the KPZ generating functional and action, which can be simply obtained following the standard Martin-Siggia-Rose-Janssen-de Dominicis  formalism~\cite{Martin73,Janssen76,Dominicis76}, and  read~\cite{Frey94}
\begin{align}
{\cal Z}[j,\tj] &= \int {\cal D} h {\cal D} \tih  \exp\left({-{\cal S}[h,\tih ] + \int_{t,\vx}\big\{ j h + \tj \tih \big\}}\right)\, ,\nonumber\\
{\cal S}[h,\tih ] &= \int_{t,\vx}\left\{\tih  \Big[\p_t h -\frac \lambda 2 (\nabla h)^2 - \nu \nabla^2 h   \Big] - D\,\tih ^2\right\}\, ,
\label{eq:actionKPZ}
\end{align}
where $\tih$ is the response field and $j,\tj$ are the sources, such that functional derivatives of ${\cal Z}$ with respect to them generate the correlation and response functions of $h$ and $\tih$.

The FRG formalism is a powerful implementation of Wilson's ideas, which is expounded in several lecture notes and reviews~\cite{Berges2002,Kopietz2010,Delamotte2012,Dupuis2021}. To achieve the progressive integration of fluctuations, one adds in the functional integral a scale-dependent weight $\exp(-\Delta \mathcal{S}_{\kappa})$, where $\kappa$ is a momentum scale, as
\begin{equation}
\label{eq:Zkappa}
    \mathcal{Z}_{\kappa}[j,\tj] = \int {\cal D} \Phi \exp \left(
        -\mathcal{S}[\Phi] - \Delta \mathcal{S}_{\kappa}[\Phi] + \int_{t,\vx} \mathcal{J}_\alpha \Phi_\alpha
    \right),
\end{equation}
denoting $\Phi=(h,\tih)$ and ${\cal J} = (j,\tj)$  the multiplet of fields and sources, indexed by $\alpha\in\llbracket 1,2\rrbracket$, and where $\Delta \mathcal{S}_{\kappa}[\Phi] \equiv \frac{1}{2}\int_{t,\vx,\vx'} \Phi_\alpha(t,\vx) \mathcal{R}_{\kappa,\alpha\beta}(|\vx-\vx'|) \Phi_\beta(t,\vx')$ is chosen quadratic in the fields. The kernel $\mathcal{R}_{\kappa}$ is called `regulator' and each of its elements  must satisfy the following  set of properties (in  Fourier space):
\begin{align}
&({\rm i}):\,\mathcal{R}_{\kappa,\alpha\beta}(\vp) \stackrel{|\vp| \ll \kappa}{\sim} \kappa^2\,,\qquad
({\rm ii}):\,\mathcal{R}_{\kappa,\alpha\beta}(\vp) \stackrel{|\vp| \gg \kappa}{\longrightarrow} 0\,,\quad\nonumber\\
  &({\rm iii}):\,\mathcal{R}_{\kappa,\alpha\beta}(\vp) \stackrel{\kappa \rightarrow \Lambda}{\sim} \Lambda^2\,,\qquad
  ({\rm iv}):\,\mathcal{R}_{\kappa,\alpha\beta}(\vp) \stackrel{\kappa \rightarrow 0}{\longrightarrow} 0\,,\label{eq:reg}
\end{align}
such that  low-momentum fluctuation modes with $|\vp|\lesssim \kappa$ are suppressed, since (i) ensures  $e^{-\Delta{\cal S}_\kappa}\simeq 0$ for these modes, while  high-momentum  fluctuation modes with $|\vp|\gtrsim \kappa$ are included and averaged over, since (ii) ensures $e^{-\Delta{\cal S}_\kappa}\simeq 1$ for these modes.
We have denoted  $\Lambda$ the (UV) scale at which the model (here~\eqref{eq:KPZ}) and its parameters are defined. In general, it can be chosen finite ($\Lambda\sim 1/a$ inverse lattice spacing) if the model is discrete, or it can be  infinite.

The central object within the FRG formalism is the effective average action $\Gamma_\kappa$, defined as the  (modified) Legendre transform of ${\cal W}_\kappa =\ln{\cal Z}_\kappa$
\begin{equation}
    \Gamma_{\kappa}[\Psi] + \Delta\mathcal{S}_{\kappa}[\Psi] =  \underset{\mathcal{J}}{\sup} \left[
     -\mathcal{W}_{\kappa}[\mathcal{J}]
        +\int_{t,\vx}\mathcal{J}_\alpha \Psi_\alpha
    \right]\,,
\end{equation}
where $\Psi=(\varphi,\tvarphi)\equiv\langle\Phi\rangle_{\cal J}$ is the multiplet of the average fields in presence of the sources.  By construction (iii), at scale $\Lambda$, all fluctuations are frozen, such that
$\Gamma_{\kappa=\Lambda}$  identifies with the KPZ action~\eqref{eq:actionKPZ}  $\Gamma_\Lambda\equiv \mathcal{S}$~\cite{Delamotte2012,Dupuis2021}. In the limit  $\kappa\to0$, all fluctuations are integrated out since the regulator vanishes (iv), 
such that $\Gamma_{\kappa=0}$ is equal to the full effective action  $\Gamma_0\equiv\Gamma$. In between these two scales, $\Gamma_{\kappa}$ provides the effective theory  at scale $\kappa$, which can be interpreted as the effective theory of the system in a volume $\kappa^{-d}$. Its evolution with the RG scale is given by
the exact Wetterich equation~\cite{Wetterich93}
\begin{equation}\label{eq:Wetterich}
    \partial_{\kappa} \Gamma_{\kappa} =
    \frac{1}{2}\tr\,\int_{\omega,\vq}
    \partial_{\kappa} \mathcal{R}_{\kappa}\,
    \cdot\,\mathcal{G}_{\kappa}\,,\qquad \mathcal{G}_{\kappa} \equiv \left(
    \Gamma_{\kappa}^{(2)} + \mathcal{R}_{\kappa}
\right)^{-1}\,
\end{equation}
where the  trace means summation over all fields and $\Gamma_{\kappa}^{(2)}$ is the Hessian of $\Gamma_\kappa$. The very idea of the Wilsonian RG is thus closely realised through this formalism.
Of course, in most situations, this functional equation cannot be solved exactly, and several approximation schemes have been devised to study it, which are non-perturbative~\cite{Dupuis2021}.
 A key aspect is that these approximation schemes are controlled and can be improved in a systematic way, such that they can provide high-accuracy results, including in some cases the best available estimates by analytical methods~\cite{Balog2019,Polsi2020}. In the following, we focus on the most useful approximations
for the KPZ equation.

\section{The KPZ fixed point}
\label{sec:KPZ}

Within the FRG formalism, symmetries play a fundamental role and can be very efficiently exploited through exact identities derived from them which are called Ward identities. In fact, it proves useful to consider extended symmetries, which are infinitesimal transformations of the fields and coordinates which do not leave the action strictly invariant, but instead yield a variation at most linear in the fields. From them can be derived more general forms of the Ward identities~\cite{Canet2022}. We first briefly list the extended symmetries for the KPZ action, and then use the associated Ward identities to constrain successive approximations of the FRG equations.

\subsection{Symmetries of the KPZ action}

The KPZ action is invariant under a shift of the height field  by an arbitrary constant $h\to h+c$. This symmetry can be extended considering instead a shift by an infinitesimal scalar function of time
\begin{equation}
\label{eq:shift}
 h(t,\vx) \to h(t,\vx) +\varepsilon(t)\,.
\end{equation}
The KPZ action is invariant under this transformation but for  the term
 \big($\int_{t,\vx}  \tih\p_t\varepsilon $\big) which is linear in the fields. This can be used to derive Ward identities, which essentially prescribe that the corresponding term in the action  $\int_{t,\vx} \tih \p_t h$ is not renormalised and the rest of the effective average action must be invariant under the transformation~\eqref{eq:shift}~\cite{Canet2011kpz}.

 In a similar manner, the statistical tilt symmetry $\vx\to \vx + \lambda \vv_0 t$,  $h\to h-\vv_0\cdot \vx$, which is simply the Galilean symmetry for the corresponding Burgers equation, can be extended to
\begin{equation}
\label{eq:Galilee}
 h(t,\vx) \to h(t,\vx) +\lambda \varepsilon(t) \cdot \nabla h - \p_t{\varepsilon}(t)\cdot \vx\,, \quad  \tih(t,\vx) \to \tih(t,\vx) +\lambda \varepsilon(t) \cdot \nabla \tih\,,
\end{equation}
where $\varepsilon(t)$ is now an infinitesimal vectorial function of time.
This extended symmetry leads to very constraining Ward identities, whose derivation can be found in~\cite{Canet2011kpz,Fontaine2023InvBurgers}. It implies in particular that $\lambda$ is not renormalised, and more generally that the operator $D_t = \p_t -\lambda \grad h\cdot \grad$, which is the covariant time derivative for the Galilean symmetry, is renormalised as a whole~\cite{Canet2011kpz}.

Lastly, in $d=1$ only, the KPZ equation is invariant under a discrete time-reversal symmetry, which corresponds to the transformation
\begin{equation}
\label{eq:TRS}
 h(t,\vx) \to -h(-t,\vx)\,\qquad  \tih(t,\vx) \to \tih(-t,\vx) +\frac \nu D \nabla^2 h(-t,\vx)\,.
\end{equation}
This symmetry entails a fluctuation-dissipation theorem relating the correlation  and response functions as $\bla \tih(t,\vx)h(t',\vxp)\bra = \frac \nu D \grad^2 \bla h(t,\vx) h(t',\vxp) \bra$ ($t'>t$). It also yields that the stationary probability distribution of height fluctuations for the KPZ equation is the same as for the EW equation in $d=1$. As a consequence, they share the same roughness exponent $\chi=1/2$. This is no longer the case in $d\neq 1$ where the time-reversal symmetry is broken.

\subsection{Simplest approximation}

The simplest approximation one can use within the FRG framework is to choose an ansatz for $\Gamma_\kappa$ which has the same form as the bare action~\eqref{eq:actionKPZ}, but with scale-dependent parameters, that is
\begin{equation}
\Gamma_\kappa[\varphi,\tvarphi] = \int_{t,\vx}\left\{\tvarphi  \Big[\mu_\kappa \p_t \varphi -\frac {\lambda_\kappa}{2} (\nabla \varphi)^2 - \nu_\kappa \nabla^2 \varphi   \Big] - D_\kappa\,\tvarphi ^2\right\}\, .
\label{eq:ansLPA}
\end{equation}
In fact,  the Ward identities associated with the shift \eqref{eq:shift} and Galilean \eqref{eq:Galilee} symmetries readily enforce at this level that  $\mu_\kappa= \mu_\Lambda=1$ and $\lambda_\kappa=\lambda_\Lambda=\lambda$.
Moreover, in $d=1$, the time-reversal symmetry \eqref{eq:TRS} further imposes
 that $\frac{\nu_\kappa}{D_\kappa} = \frac{\nu}{D}$~\cite{Canet2011kpz,Fontaine2023InvBurgers}.

Let us  mention that the regulator ${\cal R}_\kappa$ in the FRG formalism should preserve the symmetries of the action. An appropriate choice for the KPZ action is~\cite{Canet2011kpz}%,Canet2012Err}
\begin{equation}
{\cal R}_\kappa(\vq) \!=\! r\left(\frac{q^2}{\kappa^2}\right)
\left(\!\! \begin{array}{cc}
0& {\nu_\kappa} q^2\\
{\nu_\kappa} q^2 & -2 D_\kappa
\end{array}\!\!\right) \;,
\label{eq:Rk}
\end{equation}
 where one typically uses for the simplest approximation Litim cutoff function $r(x)= (1/x-1)\theta(1/x-1)$~\cite{Litim2000} with $\theta$ the Heaviside step function\footnote{Litim cutoff function has the advantage that it enables one to analytically calculate the integrals over momentum, which is generally not possible with the exponential cutoff, but the latter leads in general to more accurate results.}, while for advanced approximations one generally uses $r(x)=1/(\exp(x) -1)$.
The RG flow equations $\p_\kappa\nu_\kappa$ and $\p_\kappa D_\kappa$ of the remaining parameters  can be  calculated by projecting the exact flow equation \eqref{eq:Wetterich} for $\Gamma_\kappa$ onto the ansatz~\eqref{eq:ansLPA}. The calculation is detailed in Refs.~\cite{Fontaine2023InvBurgers,Gosteva2024}.

The next important step in the RG procedure is the rescaling.
One thus defines rescaled coordinates and fields, denoted with a hat symbol, as
\begin{equation}
\label{eq:rescaling}
 \hat{\vq} = \dfrac{\vq}{\kappa}\, , \; \hat{\omega} = \dfrac{\omega}{\kappa^2\nu_\kappa}\,  ,\; \hat{\varphi} = \sqrt{\frac{\nu_\kappa}{\kappa^{d-2}D_\kappa}} \varphi\, ,\; \hat{\tvarphi} = \sqrt{\frac{D_\kappa}{\kappa^{d+2}\nu_\kappa}}\tvarphi\,.
\end{equation}
One also defines  two effective anomalous dimensions as
\begin{equation}
\label{eq:defeta}
\eta_\kappa^\nu = -\p_s \ln \nu_\kappa\,,\qquad\eta_\kappa^D = -\p_s \ln D_\kappa\,,
\end{equation}
where $s= \ln(\kappa/\Lambda)$ is the RG `time' and $\p_s =\kappa\p_\kappa$.
 In a scaling regime, the flow reaches a fixed point, such that the effective anomalous exponents acquire fixed values, denoted with an asterisk  $\eta_\kappa^D\to \eta_*^D$ and $\eta_\kappa^\nu\to \eta_*^\nu$ when $\kappa\to 0$. It follows from \eqref{eq:defeta} that the effective noise amplitude $D_\kappa$ and effective surface tension  $\nu_\kappa$ then  behave as power-laws $D_\kappa\sim \kappa^{-\eta_*^D}$ and  $\nu_\kappa\sim \kappa^{-\eta_*^\nu}$.
From the rescaling \eqref{eq:rescaling}, one can deduce that the exponents  $\chi$ and $z$ are related  to the anomalous exponents as \cite{Kloss2012}
\begin{equation}
\label{eq:defexpo}
z=2-\eta_*^\nu \,,\qquad \chi = \dfrac 1 2(2-d +\eta_*^D-\eta_*^\nu)\,.
\end{equation}
Finally, the dimensionless effective coupling, associated with the non-linear term, is then defined as $\hat g_\kappa = \kappa^{d-2}\lambda^2 D_\kappa/\nu_\kappa^3$. Its flow equation is thus given by
\begin{equation}
\label{eq:dsg}
\p_s \hat g_\kappa = \hat g_\kappa\big(d-2 -\eta_\kappa^D+3 \eta_\kappa^\nu \big)\, ,
\end{equation}
where $\eta_\kappa^\nu$ and $\eta_\kappa^D$ are non-trivial functions of $d$ and $\hat{g}_\kappa$, which can be computed from the Wetterich equation
 and are given in Refs~\cite{Fontaine2023InvBurgers,Gosteva2024}.

 In fact, even without the explicit expressions of $\eta_\kappa^\nu$ and $\eta_\kappa^D$, one can already deduce some properties of the fixed points of~Eq.~\eqref{eq:dsg}. A fixed point is defined as a stationary solution of Eq.~\eqref{eq:dsg}, {\it i.e} it satisfies $\p_s \hat g_\kappa=0$. There are two obvious such solutions. The first one is $\hat g_* \equiv \hat g_*^{\rm \tiny EW}=0$, which corresponds to the EW solution. It is Gaussian (non interacting), and thus $\eta_*^D=\eta_*^\nu=0$ at this fixed point, which from~\eqref{eq:defexpo} yields $z=2$ and $\chi=(2-d)/2$. The second one corresponds to a finite $\hat g_*$ denoted $\hat g_*^{\rm \tiny KPZ}\neq 0$  which satisfies
 \begin{equation}
 \label{eq:defKPZ}
 (d-2 -\eta_*^D+3 \eta_*^\nu \big)=0\,.
\end{equation}
This is the implicit equation for the non-trivial KPZ fixed point. Irrespective of the value of  $\hat g_*^{\rm \tiny KPZ}$, the identity~\eqref{eq:defKPZ}  entails  from~\eqref{eq:defexpo} that $z+\chi=2$ in any dimension at this fixed point.
Finally, in $d=1$, the time-reversal symmetry further imposes $\eta_\kappa^D=\eta_\kappa^\nu\equiv \eta_\kappa$, such that the identity~\eqref{eq:defKPZ} fixes the value of  $\eta_*=1/2$ again irrespective of the corresponding value  $\hat g_*^{\rm \tiny KPZ}$. These two identities together then yield $z=3/2$. The values of the  KPZ critical exponents are thus entirely fixed by the symmetries in $d=1$, which is quite uncommon in the realm of critical phenomena and is no longer true in $d\neq 1$. Note that, even though the KPZ roughness exponent  coincides with the EW value $\chi=(2-d)/2=1/2$ in $d=1$ due to the time-reversal symmetry, the KPZ dynamical exponent is anomalous, as it significantly differs from the Gaussian EW value $z=2$ (large correction). In $d>2$, the identity~\eqref{eq:defKPZ} has two solutions, both satisfying $z+\chi=2$, which are the KPZ and the RT fixed points. The RT fixed point is somewhat peculiar since it has exact critical exponents $\chi=0$ and $z=2$ in all dimensions $d\geq 2$ where it exits~\cite{Wiese98}.

To get the actual fixed point solutions $\hat g_*$ and study their stability, one needs the explicit expressions for $\eta_\kappa^\nu$ and $\eta_\kappa^D$. Using them, one obtains~\cite{Gosteva2024}
\begin{equation}
\label{eq:dstildeg}
 \p_s \tilde{g}_s = \frac{\tilde{g}_s \left(2 d^6+4 d^5-d^4 (7 \tilde{g}_s+8)-2 d^3 (3 \tilde{g}_s+8)+d^2
   \tilde{g}_s (3 \tilde{g}_s+20)-d (\tilde{g}_s-8) \tilde{g}_s-2 \tilde{g}_s^2\right)}{2 d^5+8
   d^4+d^3 (8-3 \tilde{g}_s)-2 d^2 \tilde{g}_s+d \tilde{g}_s (\tilde{g}_s+8)-\tilde{g}_s^2}\,,
\end{equation}
where $\tilde{g}_s = v_d \hat{g}_s$ with $v_d=(2^{d-1}\pi^{d/2}\Gamma(d/2))^{-1}$ a factor related to the $d$-dimensional angular integration.

\begin{figure}[h]
\begin{center}
\tikz{
\node at (-1.5,2.7) {(a)  $d=1$};
\node at (5.2,2.7) {(b) $d=3$};
\node at (-2.,0) {\includegraphics[height=4.5cm]{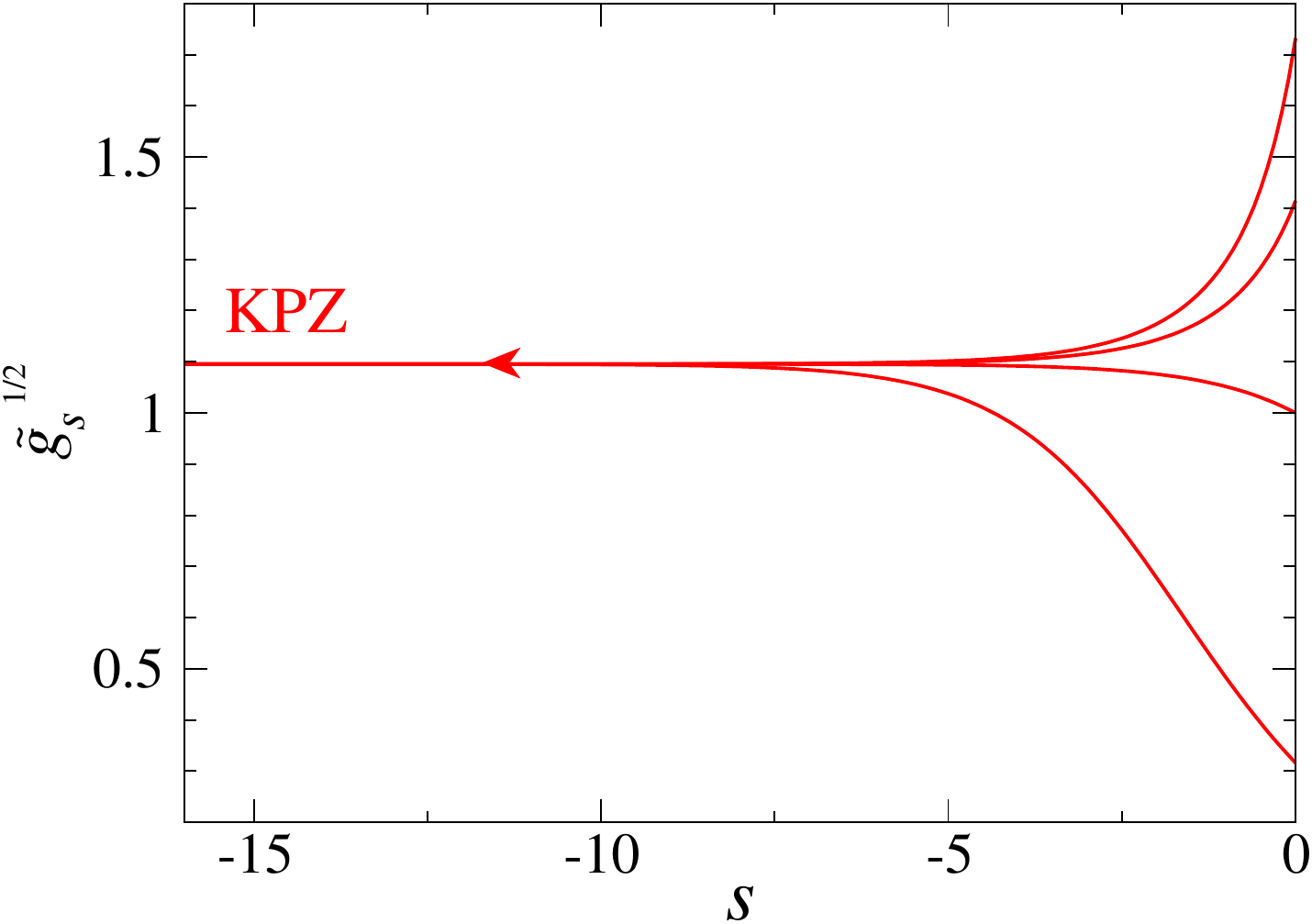}};
\node at (5,0.) {\includegraphics[height=4.5cm]{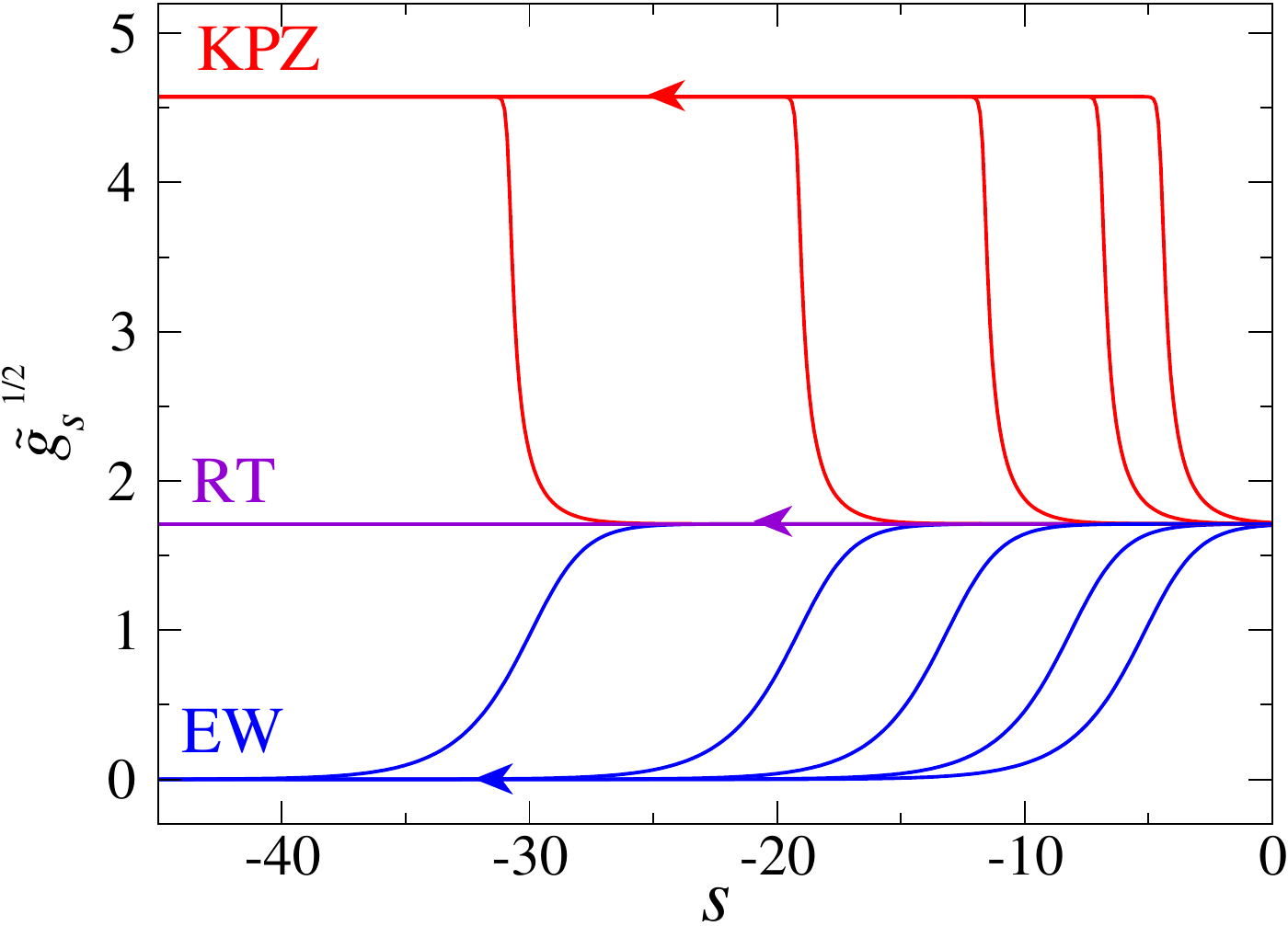}};
}
\end{center}
 \caption{Flows of $\sqrt{\tilde g_s}$ as a function of the RG time $s=\ln(\kappa/\Lambda)$ for different initial values of $\tilde g_0$ in dimension (a) $d=1$ and (b) $d=3$. The (IR) flow runs from the microscopic ($s=0$) to the macroscopic ($s\to-\infty$) scale from right to left.}
 \label{fig:flowdiag}
\end{figure}

This equation can be readily integrated numerically from the microscopic scale $s=0$ ($\kappa=\Lambda$) to the macroscopic one $s\to-\infty$ ($\kappa\to0$). Examples of flows for different initial conditions $\tilde{g}_0$ are displayed in Fig.~\ref{fig:flowdiag}.
In dimension $d=1$, the flow always reaches the same finite fixed-point value $\tilde g_*^{\rm \tiny KPZ}$ for any initial condition  $\tilde g_0$. The KPZ fixed point is thus fully attractive, which physically implies that the interface always roughens and becomes superdiffusive. In dimension $d=3$, the flow either converges to a finite value  $\tilde g_*^{\rm \tiny KPZ}> 0$ for any $\tilde g_0>\tilde g_c$, which corresponds to the KPZ fixed point, or to $\tilde g_*^{\rm \tiny EW}= 0$ for any $\tilde g_0<\tilde g_c$, which corresponds to the EW fixed point. They are both attractive. Conversely, the RT fixed point is unstable and requires fine-tuning of $\tilde g_0$ to $\tilde g_c$ to be attained.
It controls the roughening transition, while the KPZ and EW fixed points describe the rough and smooth phase respectively.
This shows that even the simplest approximation~\eqref{eq:ansLPA} allows one to capture the KPZ fixed point in any $d$. It  thus already goes beyond the all-order perturbative result, and  demonstrates  the full ability of the FRG approach to tackle strongly-coupled non-perturbative problems.

\begin{figure}[h]
\begin{center}
\includegraphics[height=5cm]{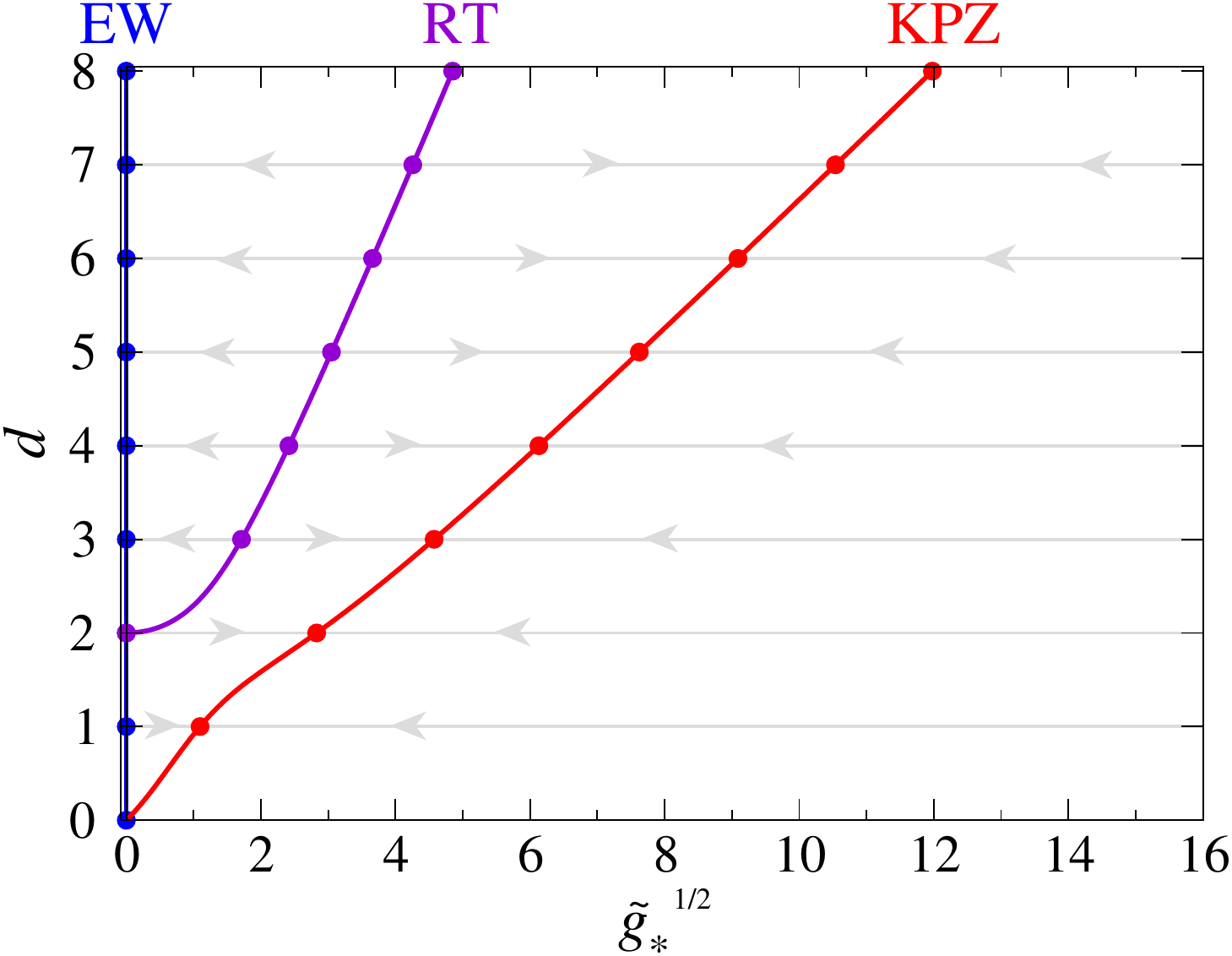}
\end{center}
 \caption{Flow diagram of the KPZ equation:  fixed point values $\sqrt{\tilde g_*}$  while varying the dimension $d$ in the vertical axis. The dots represent the  fixed points, and the grey lines with arrows indicate the RG flows (from $s=0$ to $s\to-\infty$) which show whether these fixed points are stable are unstable (in the IR flow).}
 \label{fig:phasediag}
\end{figure}

All the fixed point solutions, both stable and unstable, can be determined from~\eqref{eq:dstildeg} in any dimension. One can construct from them the flow (and phase) diagram of the KPZ equation, which is represented in Fig.~\ref{fig:phasediag}.
It shows that the KPZ fixed point is always attractive in any $d$, while the RT fixed point is always repulsive. The EW fixed point changes stability in $d=2$, from unstable in $d\leq 2$ to stable in $d>2$, beyond which it describes a smooth phase  which is attained for  small non-linearity $\tilde g_0 \leq \tilde g_c$. Let us notice that a very efficient way to evidence the unstable fixed points, rather than fine-tuning the initial condition, is to reverse the flow, {\it i.e} change $s\to-s$,  and run it from some very large (negative) value of $s$ to $s\to 0$. Indeed, the change of sign inverses the stability properties, such that the  unstable fixed points (in the IR flow,which runs from $s=0$ to $s\to -\infty$) become  stable (in the UV flow, which runs from $s=-\infty$ to $s\to 0$) and reciprocally. For this reason, we call the fixed points in the following IR fixed points  when they are stable in the IR flow, and UV fixed points when they are stable in the UV flow.

Even though this simplest approximation already provides a full qualitative picture, it yields very poor values for the critical exponents in $d>1$~\cite{Fontaine2023InvBurgers}.
 To obtain quantitative results, one needs to resort to more advanced approximations, as presented in the next section.

\subsection{Advanced approximations}

The next levels of approximation consist in upgrading the scale-dependent parameters $\mu_\kappa$, $\nu_\kappa$ and $D_\kappa$ to whole functions of the momentum and frequency, {\it eg}. $\nu_\kappa \to \nu_\kappa(\omega,\vp)$, and similarly for
 the others. In real space, it means that they become functions of the derivative operators $\grad$ and $\p_t$, or more appropriately  $D_t = \p_t -\lambda \grad \varphi\cdot \grad$, which has the property to automatically preserve the Galilean symmetry~\cite{Canet2011kpz}.
The general ansatz for the effective average action, enforcing the constraints from the shift~\eqref{eq:shift} and Galilean~\eqref{eq:Galilee} symmetries, reads
\begin{equation}
\Gamma_\kappa[\varphi,\tvarphi]= \int_{t,\vx}
\left\{ \tvarphi \left[\mu_\kappa(D_t,\grad) \left(\p_t\varphi - \frac{\lambda}{2} (\grad \varphi)^2\right)
 -   \nu_\kappa(D_t,\grad)  \grad^2 \varphi\right] - D_\kappa(D_t,\grad) \tvarphi^2 \right\}\,.
\label{eq:ansSO}
\end{equation}
In $d=1$, the time-reversal symmetry further imposes that $\mu_\kappa(D_t,\grad)\equiv 1$ and $D_\kappa(D_t,\grad)=\frac D \nu \nu_\kappa(D_t,\grad)$, such that one is left with a single renormalisation function.

In fact, three successive levels of approximation, of increasing accuracy, have been considered. Generically denoting the renormalisation functions $f_\kappa$, with $f\equiv \mu, \nu$ or $D$, these approximations consist in assuming a functional dependence on momentum only $f_\kappa(\grad)$
(referred to as LO (Leading Order) approximation~\cite{Canet2010}), on momentum and frequency $f_\kappa(\p_t,\grad)$ (referred to as NLO (Next-to-Leading Order) approximation~\cite{Kloss2012}), or on momentum, frequency, and fields through the covariant derivative $f_\kappa(D_t,\grad)$ (referred to as SO (Second Order) approximation, since it is the most general ansatz at quadratic order in the response field~\cite{Canet2011kpz}).

\begin{figure}[h]
\begin{center}
\tikz{
\node at (-1.8,2.7) {(a)  scaling function $F_{\rm KPZ}(pt^{2/3})$};
\node at (5.5,2.7) {(b) scaling function $\bar{F}_{\rm KPZ}(x/t^{2/3})$};
\node at (-2.,0) {\includegraphics[height=4.5cm]{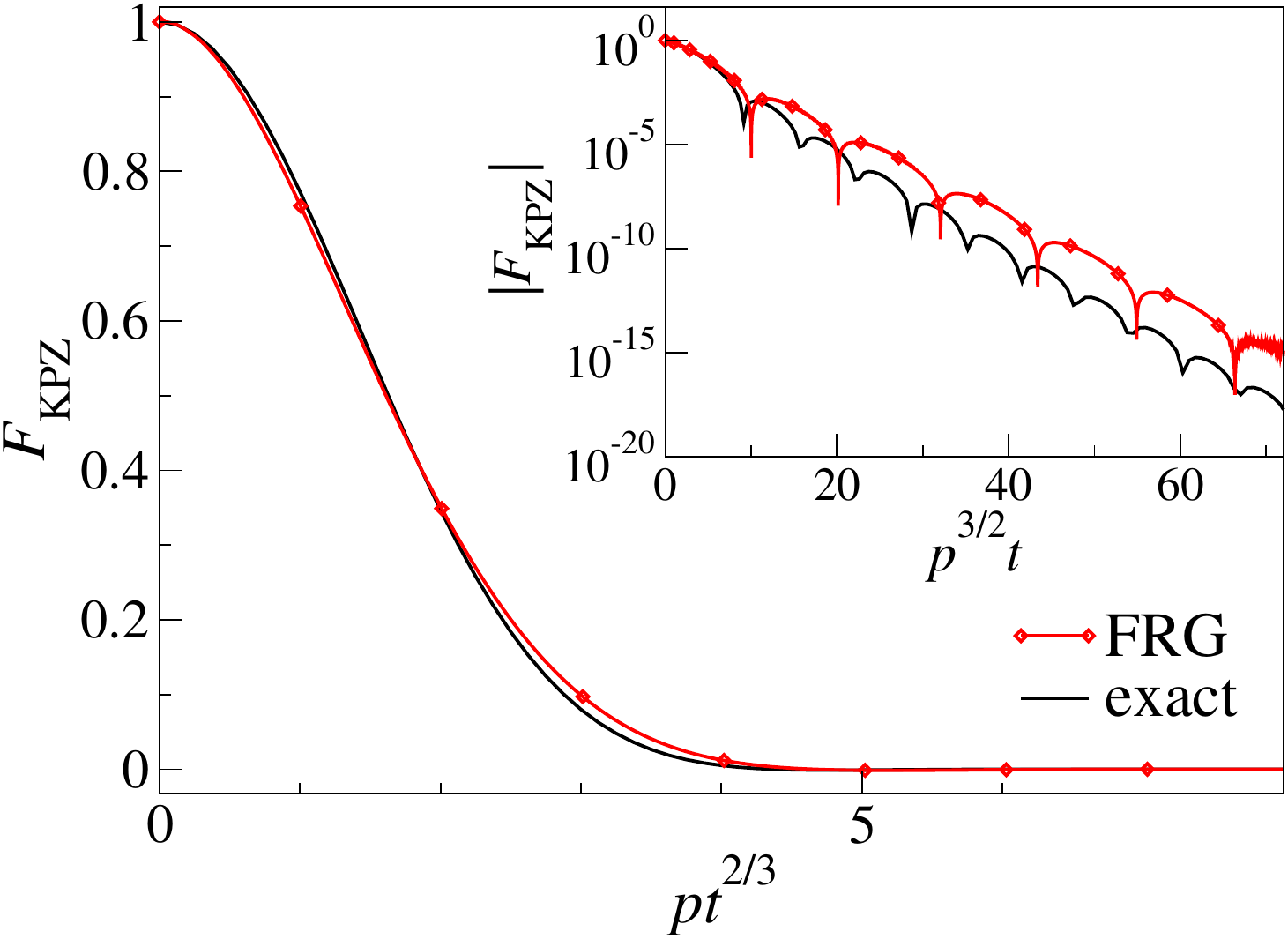}};
\node at (5,0.) {\includegraphics[height=4.5cm]{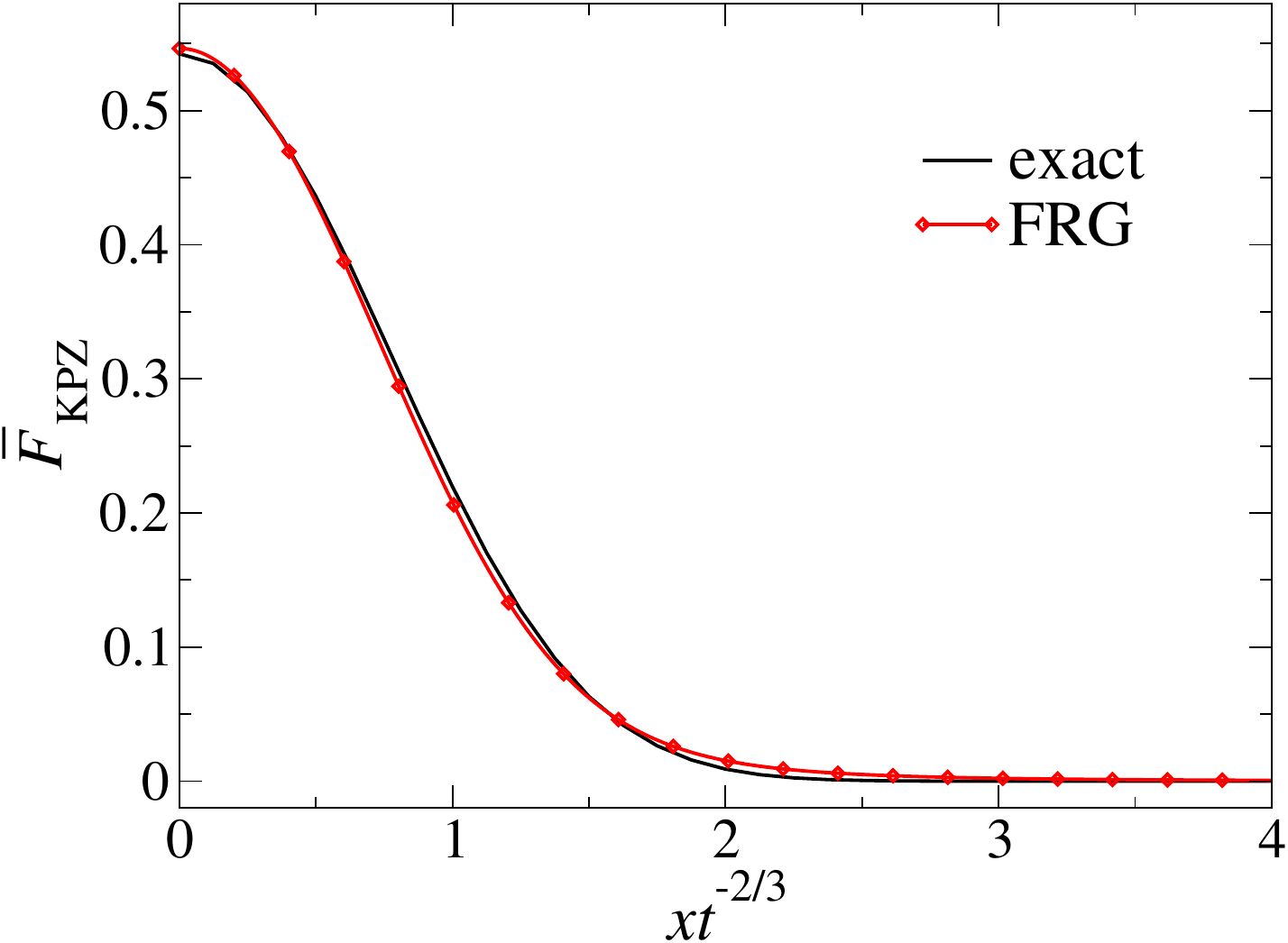}};
}
\end{center}
 \caption{Scaling functions in $d=1$ associated with the correlation function (a) in Fourier space $C(t,p)$ and (b) in real space $C(t,x)$, from the FRG calculation~\cite{Canet2011kpz} and from the exact result of Ref.~\cite{Praehofer04}. There are no adjustable parameters.}
 \label{fig:scaling-function-d1}
\end{figure}

These approximations can be tested in $d=1$ by comparing with the available exact results.
In particular, the scaling function $\bar{F}_{\rm KPZ}(xt^{-2/3})$   associated with the correlation function $C(t,\vx)$~\eqref{eq:correlation} was calculated exactly for the stationary subclass in Ref.~\cite{Praehofer04}, together with the one ${F}_{\rm KPZ}(pt^{2/3})$ associated with its Fourier transform in space $C(t,\vp)$. Both were computed within the SO approximation of the FRG~\cite{Canet2011kpz}. The results are displayed in Fig.~\ref{fig:scaling-function-d1}, which shows a nearly perfect agreement with the exact results, without any adjustable parameters. Let us point out that the exact function ${F}_{\rm KPZ}(pt^{2/3})$ is known to exhibit a negative dip, followed by a stretched exponential tail with superimposed tiny oscillations. As shown in the inset of Fig.~\ref{fig:scaling-function-d1}(a), these very fine features are reproduced by the FRG calculation down to extremely small magnitudes.

\begin{figure}[h]
\begin{center}
\tikz{
\node at (-1.8,2.7) {(a)  scaling function $\mathring{F}_{\rm KPZ}(\omega/p^z)$};
\node at (5.5,2.7) {(b) scaling function $\mathring{H}_{\rm KPZ}(\omega/p^z)$};
\node at (-2.,0) {\includegraphics[height=4.5cm]{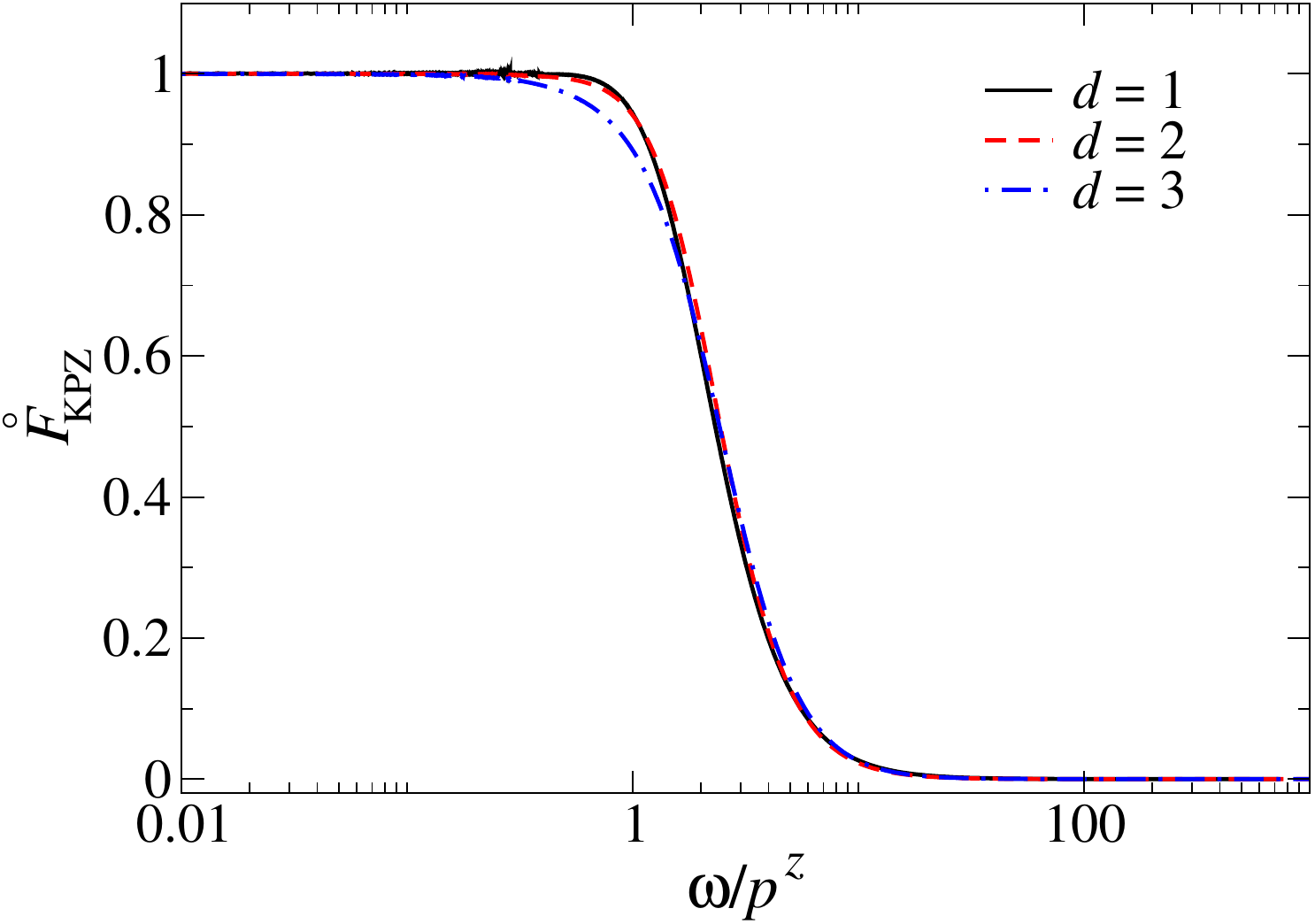}};
\node at (5,0.) {\includegraphics[height=4.5cm]{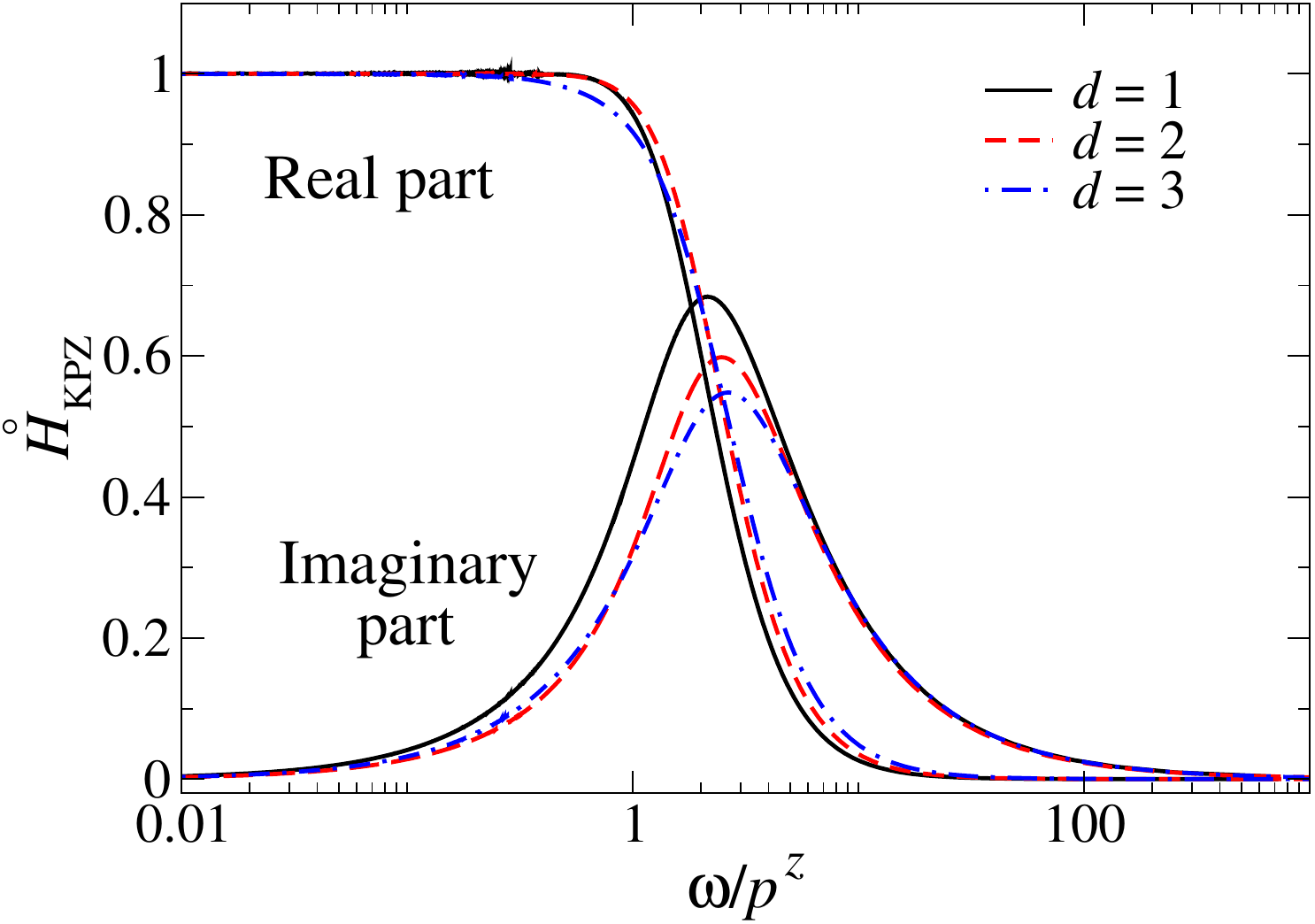}};
}
\end{center}
 \caption{Scaling functions from the FRG calculation in $d=1,2,3$: (a) $\mathring{F}_{\rm KPZ}$ associated with the correlation function and (b) $\mathring{H}_{\rm KPZ}$ associated with the response function~\cite{Kloss2012}.}
 \label{fig:scaling-functions-d}
\end{figure}

The characterisation of the KPZ fixed point has been extended to higher dimensions within the NLO approximation~\cite{Kloss2012}. The results for the scaling function $\mathring{F}(\omega/p^z)$ associated with the Fourier transform in space and time  $C(\omega,\vp)$ of the correlation function in $d=1,2,3$ are displayed in Fig.~\ref{fig:scaling-functions-d}(a). In $d\neq 1$, the response function is no longer related to the correlation function by a fluctuation-dissipation theorem and is an independent function. It has also been determined using FRG and the associated scaling function $\mathring{H}(\omega/p^z)$ (which has both real and imaginary parts) is shown in Fig.~\ref{fig:scaling-functions-d}(b). Subsequent numerical simulations have very accurately confirmed these predictions~\cite{Halpin-Healy13}.%,Halpin-Healy13Err}.

As mentioned in the introduction, the FRG has also been  used to study the effect of non-delta correlation in the noise~\eqref{eq:noise}, both spatial correlations,  long-range~\cite{Kloss2014a} and of finite range~\cite{Mathey2017}, and long-range temporal correlations~\cite{Squizzato2019}.
Another interesting ingredient is the presence of spatial anisotropy, where the full non-perturbative phase diagram has been established using FRG~\cite{Kloss2014b}. This situation has recently attracted a renewed attention in the context of  driven-dissipative Bose-Einstein condensates,  realised in particular with exciton-polaritons. This system can be described by a generalised stochastic Gross-Pitaevskii equation, which is equivalent to the complex Ginzburg-Landau equation (Eq.~\eqref{eq:CGLE_c1c2} of Sec.~\ref{sec:CGLE}). In a certain regime, the phase of the condensate wavefunction can be mapped to a KPZ equation (or Kuramoto-Sivashinsky equation Eq.~\eqref{eq:KS}), leaving deep imprints on the coherence properties of the condensate. While the KPZ universal properties have been observed in experiments in a one-dimensional exciton-polariton system~\cite{Fontaine2022Nat}, the existence of the KPZ phase in $d=2$ has been highly debated. It has been specifically argued  that anisotropy could favour the emergence of the KPZ phase, which renders this ingredient particularly salient.   We do not delve on these applications here, and now focus on the inviscid limit of the KPZ equation.

\section{The Inviscid Burgers fixed point}
\label{sec:IB}

We first show, within the simplest approximation of the FRG, the existence of a new fixed point of the KPZ equation, unveiled in~\cite{Fontaine2023InvBurgers}, and which had been missed so far. We then turn to more advanced approximations to quantitatively characterise it.

\subsection{Simplest approximation}

As emphasised in the introduction, an unexpected scaling regime, characterised by a dynamical exponent $z=1$ has been observed in $d=1$ in several numerical simulations of the KPZ or Burgers equation, when taking the limit of vanishing  surface tension or viscosity $\nu$. This scaling regime should be controlled by a new fixed point, different from the known KPZ, EW or RT ones. However, taking the inviscid limit is equivalent to taking the limit of infinite effective non-linearity $g=\lambda^2 D/\nu^3\to\infty$. This limit is obviously not accessible through a perturbative expansion in $g$, such that the non-perturbative side of the KPZ equation turns out to also arise in $d=1$ !

To study a fixed point located at infinity, it is convenient to change variables, defining the new coupling $\hat w_s = \hat g_s/(1+\hat g_s)$, such that a fixed point $\hat g_*=\infty$
  is converted to a finite value $\hat w_*=1$. The flow equation for $\hat w_s$ can be straightforwardly derived from the flow of $\hat g_s$ \eqref{eq:dsg} and reads~\cite{Gosteva2024}
 \begin{equation}
\label{eq:dsw}
\p_s \hat w_s = \hat w_s(1-\hat w_s)\big(d-2 -\eta_s^D+3 \eta_s^\nu \big)\, .
\end{equation}
Besides the fixed point solutions already discussed in Sec.~\ref{sec:KPZ}, this equation reveals the existence of an additional fixed-point solution $\hat w_*=1$, that is  $\hat g_*=\infty$, which has been named the inviscid Burgers (IB) fixed point~\cite{Fontaine2023InvBurgers}. This fixed point is a UV fixed point, {\it ie} it is unstable,  in all dimensions.

\begin{figure}[h]
\begin{center}
\includegraphics[height=5cm]{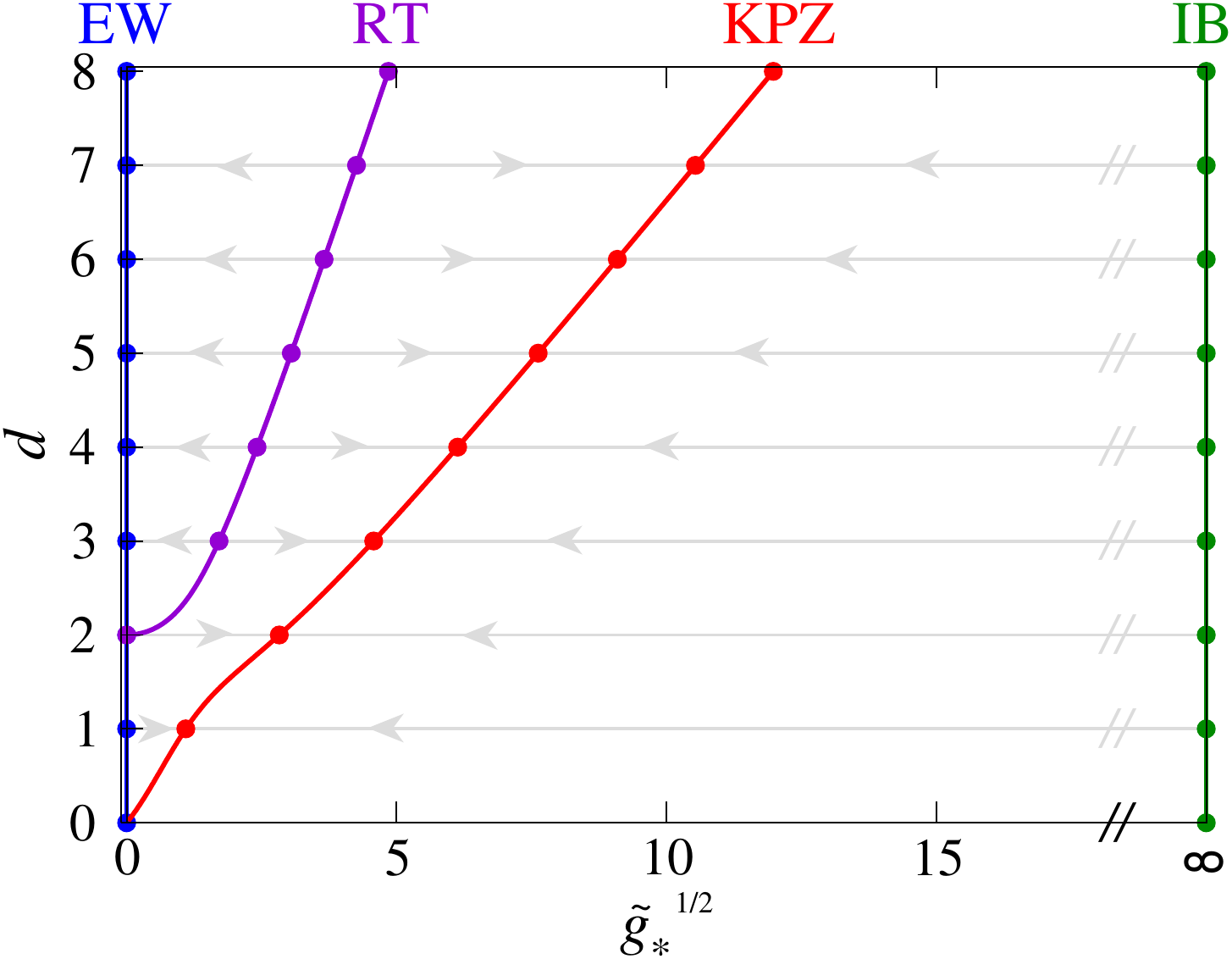}
\end{center}
 \caption{Same as Fig.~\ref{fig:phasediag}, but featuring the additional IB fixed point, which is a UV fixed point, {\it ie} always repulsive.}
 \label{fig:phasediagIB}
\end{figure}

The flow equation~\eqref{eq:dsw} can again be very easily integrated numerically, using the trick to revert the direction of the RG flow $s\to -s$ to evidence UV fixed points. The now complete flow diagram of the KPZ equation is depicted on Fig.~\ref{fig:phasediagIB}, showing the uncovered IB fixed point.

Since the KPZ fixed point is always attractive, one could wonder why an unstable  fixed point should play any role. Of course, it does so when $\nu=0$ exactly. More interestingly, even at finite but small viscosity, the IB fixed point turns out to control the large-momentum behaviour of correlation functions. In fact, the same occurs for finite but small non-linearity $\lambda$ with the EW fixed point, which is also a UV fixed point and  controls the large-momentum behaviour of correlation functions, in the opposite regime ($g\to 0$) compared with the IB fixed point ($g\to\infty$). This is developed in the next section.

\subsection{Advanced approximations}
\label{sec:IBmomenta}

The IB fixed point has been quantitatively characterised within the NLO approximation in $d=1$ in Refs.~\cite{Fontaine2023InvBurgers,Gosteva2025}, which first established that indeed  $z\simeq 1$ at the IB fixed point.
Moreover, to simultaneously resolved small momenta (controlled by the IR fixed point) and large momenta (controlled by the UV fixed points), a specific numerical scheme, based on a two-grid approach, has been devised in Ref.~\cite{Gosteva2025}.
It has been used to compute the full correlation function $C(\omega,p)$ on an extended range of $p$.

\begin{figure}[h]
\begin{center}
\tikz{
\node at (-4,0) {\includegraphics[height=5cm]{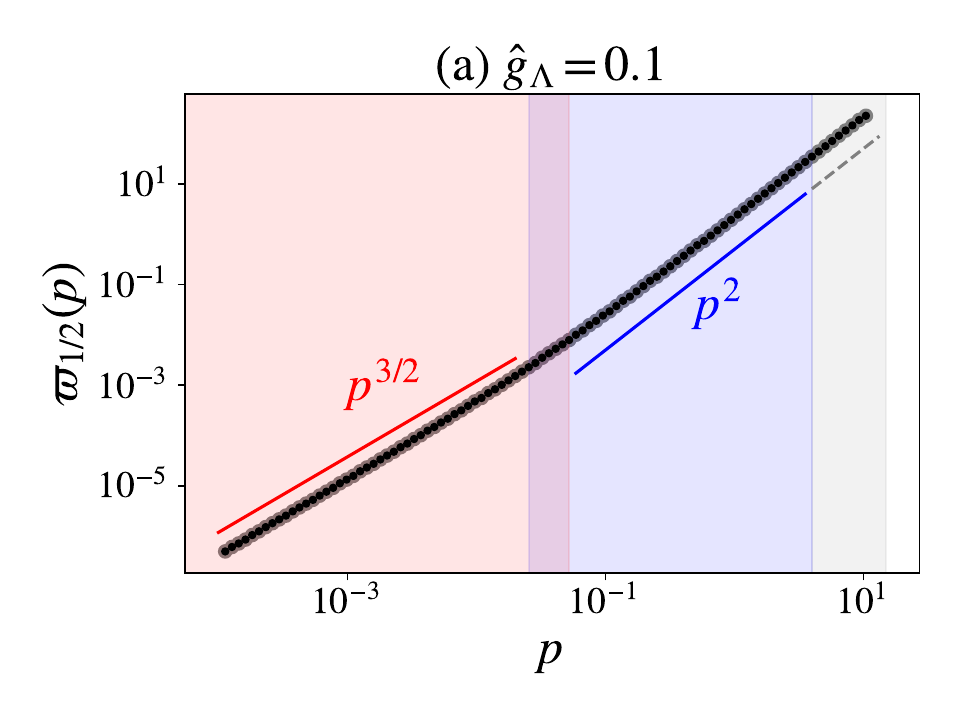}};
\node at (3,0.1) {\includegraphics[height=5cm]{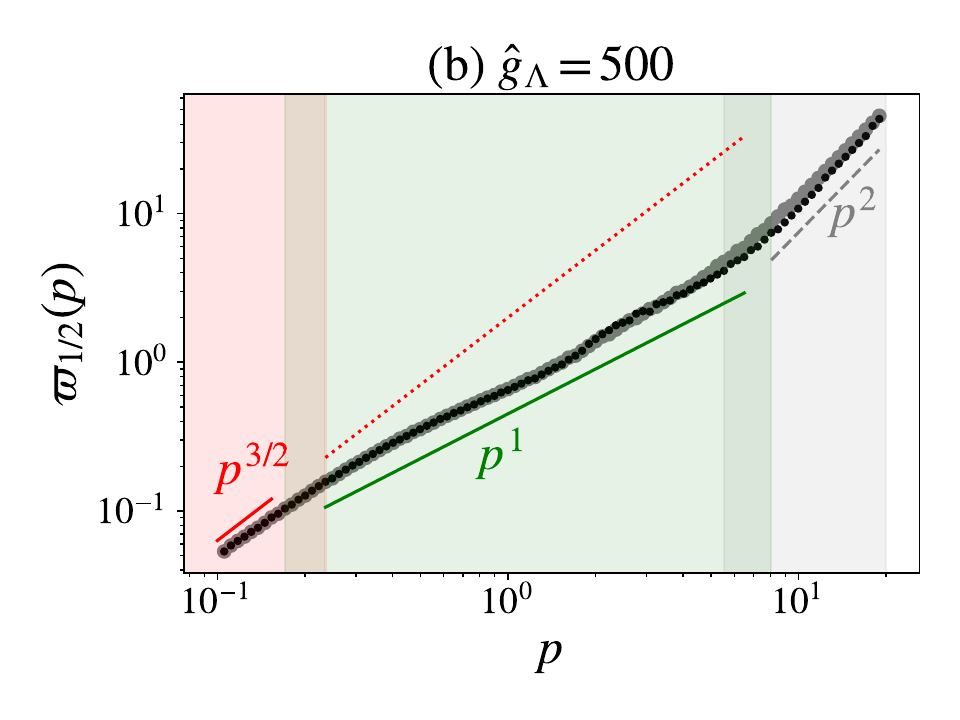}};
\node at (0,-2.6) {(c) scheme of the fixed point structure};
\node at (0,-3.5) {\includegraphics[width=6cm]{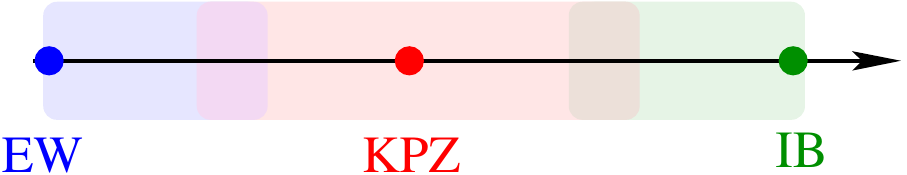}};
}
\end{center}
 \caption{Half decay frequency $\varpi_{1/2}(p)$ for (a) $\hat{g}_\Lambda=0.1$ and (b) $\hat{g}_\Lambda=500$, which lie respectively below and above the KPZ fixed point value $\hat g_*^{\rm \tiny KPZ}$. (c) Scheme of the fixed point structure of the KPZ equation in $d=1$. The different shades symbolise the regions in momenta which are controlled by each fixed point, and correspond to the same shades in (a) and (b). The grey region in (a) and (b) at highest momenta is non-universal, and merely reflects the initial condition. Figures taken from~\cite{Gosteva2025}.}
 \label{fig:omega-onehalf}
\end{figure}

 To evidence dynamical scaling regimes, a useful  probe is the half decay frequency, defined for each momentum $p$ as the frequency for which the correlation has decayed to half its value at zero frequency, that is $C(\varpi_{1/2}(p),p)=C(0,p)/2$. In a scaling regime, this quantity is expected to behave as a power-law  characterised by  the dynamical exponent $z$ as $\varpi_{1/2}(p)\sim p^z$.
 The half decay frequency obtained for both a  small and a large value of $\hat g_\Lambda$ is shown in Fig.~\ref{fig:omega-onehalf}, together with a scheme of the fixed point structure of the KPZ equation in $d=1$.
The shaded regions in this scheme symbolise the range of momenta controlled by the fixed points: IR momenta for the IR KPZ fixed point, and UV momenta for the UV EW and IB fixed points. If the initial  $\hat{g}_\Lambda$ lies below the KPZ fixed point value  $\hat g_*^{\rm \tiny KPZ}$, then the UV momenta are controlled by the EW fixed point. This is indeed what is observed in Fig.~\ref{fig:omega-onehalf}(a)  for $\varpi_{1/2}(p)$ which shows at small momenta (large distances) the expected $z=3/2$ KPZ scaling, and at large momenta the diffusive $z=2$ EW scaling.
However, if the initial  $\hat{g}_\Lambda$ lies above $\hat g_*^{\rm \tiny KPZ}$, then the UV momenta are controlled instead by the IB fixed point, as shown in Fig.~\ref{fig:omega-onehalf}(b). Here, the half decay frequency follows on a wide range the $z=1$ IB scaling, while the small momenta still display the KPZ scaling as expected. The range of the IB scaling was shown to increase with  $\hat{g}_\Lambda$, and to disappear when  $\hat{g}_\Lambda\sim \hat g_*^{\rm \tiny KPZ}$~\cite{Gosteva2025}. Let us notice that the mathematical studies are built on the EW-KPZ trajectory, and have not explored yet the other side of the KPZ fixed point, which would be a nice endeavour.

\subsection{Comparison with direct numerical simulations}

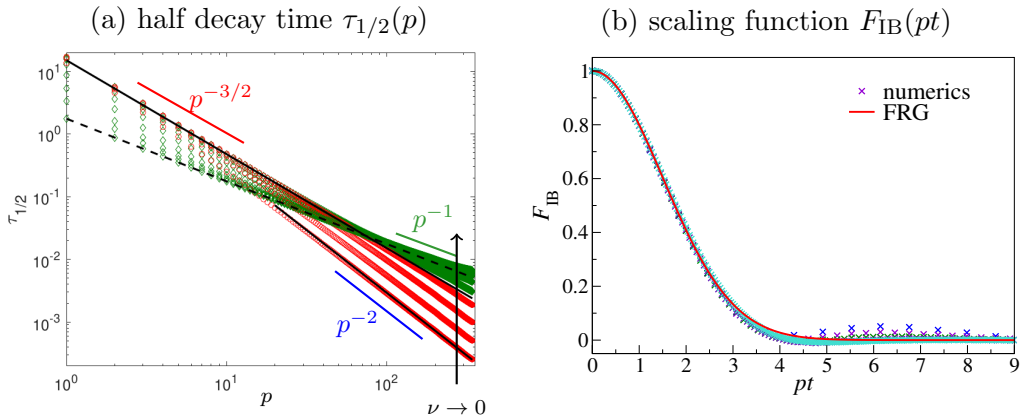
\begin{figure}[h]
\begin{center}
\tikz{
\node at (-3.8,2.8) {(a) half decay time $\tau_{1/2}(p)$};
\node at (3,2.8) {(b) scaling function $F_{\rm IB}(pt)$};
\node at (-4,0) {\includegraphics[height=5cm]{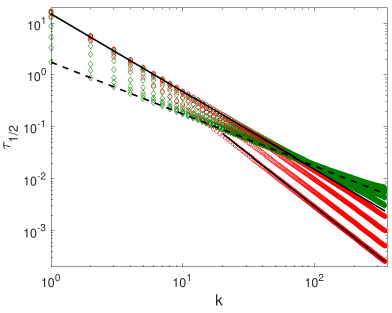}};
\node at (3,0.1) {\includegraphics[height=4.5cm]{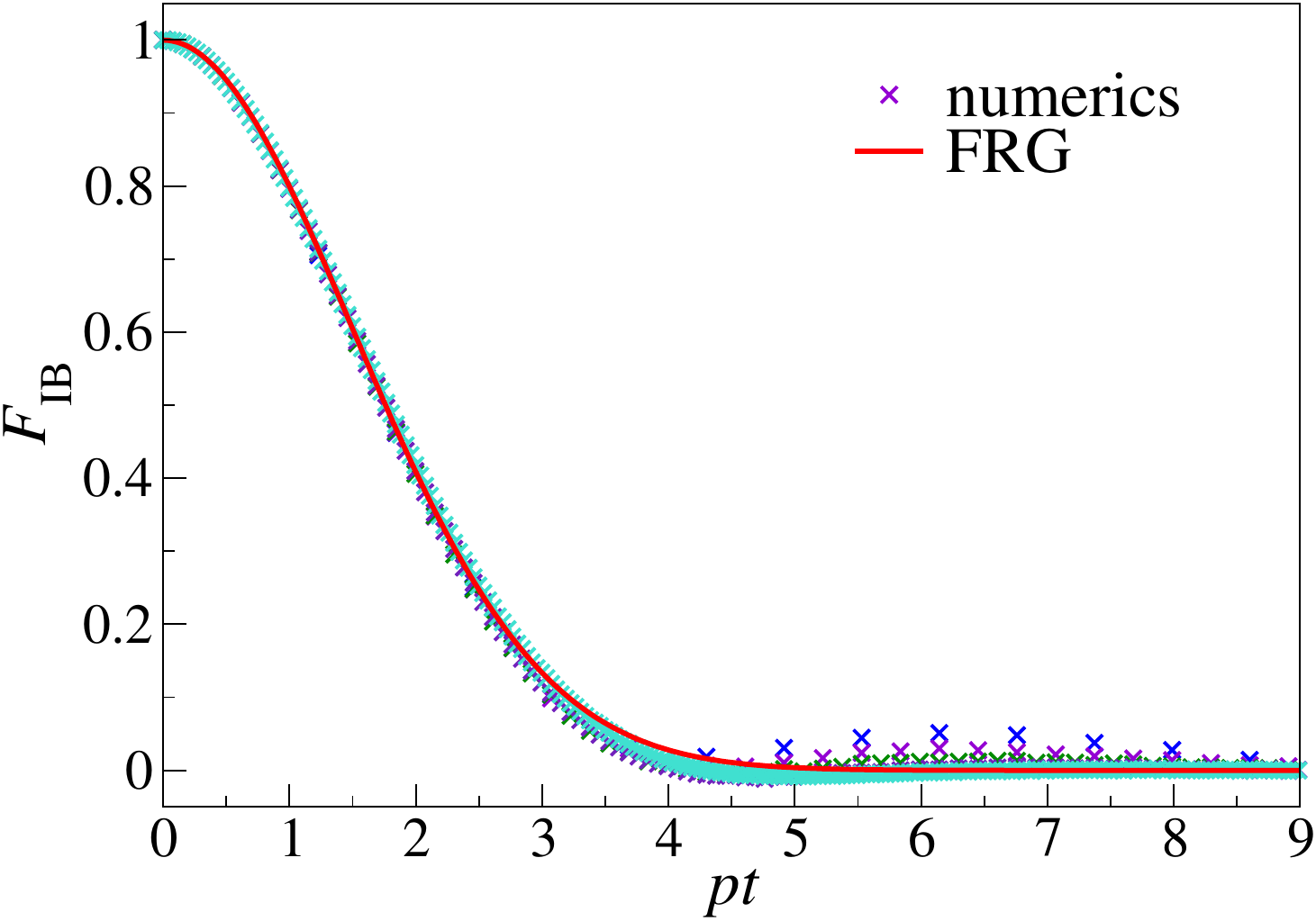}};
\draw[-,red,thick] (-5.4,2) -- (-4.,1.2);
\node at (-4.3,1.8) {\small \color{red} $p^{-3/2}$};
\draw[-,blue,thick] (-2.8,-0.5) -- (-1.65,-1.4);
\node at (-2.5,-1.2) {\small \color{blue} $p^{-2}$};
\draw[-,medgreen,thick] (-2.,0.) -- (-1.2,-0.3);
\node at (-1.5,0.2) {\small \color{medgreen} $p^{-1}$};
\fill[white] (-4,-2.4) rectangle (-3.5,-2.);
\node at (-3.7,-2.2) {\scriptsize $p$};
\draw[->,black,thick] (-1.2,-2) -- (-1.2,0.);
\node at (-1.2,-2.3) {\scriptsize $\nu\to 0$};
}
\end{center}
 \caption{(a) Half decay time for different viscosities from the numerical simulations of Ref.~\cite{Brachet2022} (Figure adapted from this reference). (b) Scaling function at the IB fixed point: the plain line shows the exact asymptotic solution~\eqref{eq:solC} of the FRG equations, while the symbols are  numerical data from Ref.~\cite{Brachet2022}.}
 \label{fig:scaling-function-IB}
\end{figure}

The FRG analysis fully explains the observation from the numerics. Indeed, in Ref.~\cite{Brachet2022}, the equivalent half decay time defined from $C(t,p)$ as  $C(\tau_{1/2}(p),p)=C(0,p)/2$ has been computed. The expected scaling behaviour is now $\tau_{1/2}(p)\sim p^{-z}$. The result   is shown in Fig.~\ref{fig:scaling-function-IB}(a) for different viscosities.
At small momenta, the KPZ scaling $z=3/2$ is always observed, apart when $\nu=0$ which corresponds to the dots on top of the black dashed line. At large momenta, the scaling crosses over from $z=2$ at large $\nu$ to $z=1$ at small $\nu$, in complete agreement with the FRG prediction.

To complete the quantitative comparison, one can focus on the IB scaling function $F_{\rm IB}$ associated with the correlation function $C(t,p)$ when $\nu\to0$. It  has also been determined in Ref.~\cite{Fontaine2023InvBurgers} within the NLO approximation, where it was found to accurately compare with the one obtained from the numerical data of Ref.~\cite{Brachet2022}.

In fact, in this case, the exact asymptotic form of the scaling function can be determined directly from the exact FRG flow equations~\eqref{eq:Wetterich}, without relying on an ansatz such as~\eqref{eq:ansLPA} or~\eqref{eq:ansSO} like in the approximations presented so far. This is because for the IB fixed point one is interested in the large-momentum behaviour. It turns out that in the limit of large momenta, the Wetterich flow equation~\eqref{eq:Wetterich} can be closed exactly using the Ward identities related to the extended symmetries~\eqref{eq:shift} and~\eqref{eq:Galilee} of the KPZ equation, without other approximations than the large-momentum limit~\cite{Fontaine2023InvBurgers,Gosteva2024}. This remarkable exact closure of the FRG equations at large momenta is not restricted to the KPZ equation, it has also been realised in other contexts such as the Navier-Stokes equation~\cite{Tarpin2018,Canet2022} or passive scalar turbulence~\cite{Pagani2021}. The analytical fixed-point solution for the correlation function from these exact asymptotic FRG equations reads
\begin{equation}\label{eq:solC}
    C(t,\vp) =  C(0,\vp){\times}
    \begin{cases}
        \exp\left( - \mu_0 (|\vp|t)^2 \right),\quad t\ll \tau_c
        \\
        \exp\left( - \mu_{\infty} |\vp|^2 |t| \right),\quad t\gg \tau_c
    \end{cases}
\end{equation}
where $\mu_0$, $\mu_{\infty}$ are non-universal constants and $\tau_c$ is a characteristic time scale.  At small time delays, this is a scaling form in the variable $|\vp|^z t \equiv |\vp| t$,  which  demonstrates that $z=1$, and provides the asymptotic form of the associated scaling function, which is simply a Gaussian. It is compared in Fig.~\ref{fig:scaling-function-IB}(b) with the scaling function obtained from the numerical data, and the two  perfectly match at small times. In fact, the solution~\eqref{eq:solC} holds in all dimensions, and thus shows that the value $z=1$ is super-universal in the sense that it does not depend on the dimension. Let us notice that this is in agreement with a result from calculations based on the replica method applied to  the equivalent problem of directed polymers in random media, which found that, in the limit $d\to\infty$ where the replica method becomes exact, the dynamical exponent is $z=1$  when $\nu\to 0$~\cite{Bouchaud1995}. This is also in accordance with results for the deterministic and spectrally-truncated inviscid Burgers equation in $d=1$~\cite{Majda2000}.

Interestingly, the solution~\eqref{eq:solC} also predicts a crossover to another regime at large time delays, characterised by a much slower exponential decay. 
Hints of this regime can be observed in the numerical data of Fig.~\ref{fig:scaling-function-IB}(b) as their lack of collapse in the variable $pt$ at large time, which indeed indicates a change of scaling. However, these data do not have enough resolution at large time delays to make a quantitative comparison with the solution~\eqref{eq:solC}. Note that a similar crossover has  also been predicted for the Navier-Stokes equation and passive scalar turbulence. In this context, the two temporal regimes can be related to the two different behaviours of the mean-square displacement of a Lagrangian particle in a turbulent flow, which behaves ballistically at small time and diffusively at large times, as shown in the seminal paper by Taylor~\cite{Taylor22}. This interpretation is expounded in Ref.~\cite{Gorbunova2021}.
The large time regime could be evidenced and quantitatively confirmed in  direct numerical simulations of such turbulent flows~\cite{Gorbunova2021scalar,Gosteva2025Euler}.

\subsection{The realm of the IB fixed point}
\label{sec:CGLE}

To finish this overview, let us emphasise that in fact the relevance of the IB fixed point is much wider than the sole inviscid limit of the KPZ equation. Indeed, it generically appears in broader contexts, in particular the one-dimensional complex Ginzburg-Landau (CLG) equation~\cite{Aranson2002} which describes the dynamics of a complex scalar order parameter $\psi$ as
\begin{equation}
    i\partial_t \psi = i\psi + (c_2 - i)|\psi|^2 \psi - (c_1 - i) \partial_x^2 \psi\,,
    \label{eq:CGLE_c1c2}
\end{equation}
where $c_1$, $c_2$ are dimensionless real coefficients,
and the Kuramoto-Sivashinsky (KS) equation~\cite{Kuramoto1978, Sivashinsky1977} which describes the deterministic and chaotic behaviour of a height field according to
\begin{equation}
    \partial_t h = \nu \partial_x^2 h  + \tau \partial_x^4  h + \frac{\lambda}{2} (\partial_x h)^2\,.
\label{eq:KS}
\end{equation}
This is not the place to review these two equations, which are  whole worlds in themselves. The interesting point here is that the CLG equation exhibits in a certain $(c_1,c_2)$ region a phase called phase turbulence, in which the dynamics of the phase of the complex field $\psi$ can be mapped onto the KS equation.
 The KS equation is characterised by a negative microscopic viscosity $\nu\equiv\nu_\Lambda<0$, and hence requires a fourth order gradient with $\tau>0$ for global stability. The negative viscosity generates chaos, since it renders a range of momenta linearly unstable. The resulting chaoting dynamics is known to belong at large space and time scales to the KPZ universality class~\cite{Cuerno1995, Ueno2005}, although it requires huge system sizes and times to actually be observed in numerical simulations~\cite{Roy2020}. Otherwise the large-scale behaviour exhibits EW scaling instead.

\begin{figure}[h]
\begin{center}
\tikz{
\node at (-4,2.9) {(a) correlation function $C(t,k)$};
\node at (3.3,2.9) {(b) half decay time $\tau_{1/2}(k)$};
\node at (-4,0) {\includegraphics[height=5cm]{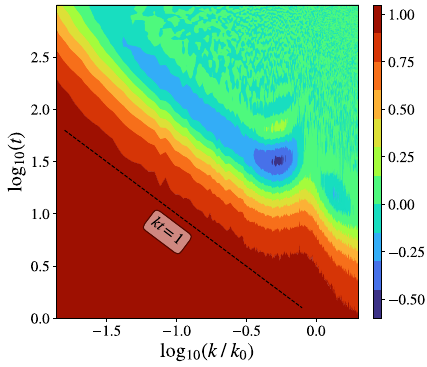}};
\node at (3,0.1) {\includegraphics[height=5cm]{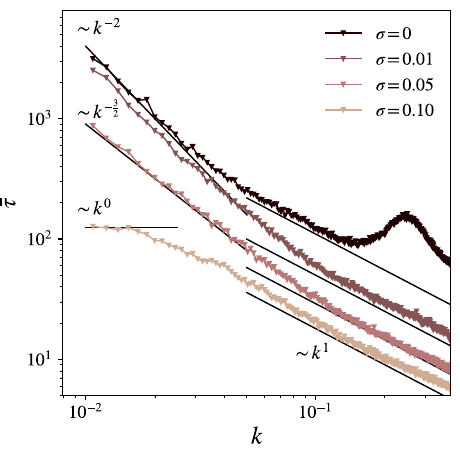}};
}
\end{center}
 \caption{Results from numerical simulations of the complex Ginzburg-Landau equation. (a) Map of the correlation function $C_{\psi\psi}(t,k)$ in the $(k,t)$ plan  in logarithmic  scales. The level lines demonstrates that the scaling is $k^zt\equiv kt$, {\it ie} $z=1$. (b) Half decay time $\tau$ as a function of $k$ for different noise strengths $\sigma$.  Figures taken from~\cite{Vercesi2024}.}
 \label{fig:CGLE}
\end{figure}

From the emergence of the KPZ universality class one infers that at large scales, an effective noise must be generated, and the effective viscosity must change sign to become positive $\nu_{\rm eff}>0$ (the fourth order gradient term becoming then irrelevant). Hence, in the process, the effective viscosity must necessarily vanish.
According to the analysis presented in Sec.~\ref{sec:IBmomenta}, this implies that there must be a whole intermediate range of momenta, corresponding to the region when  $\nu_\kappa\simeq 0$, which is controlled by the IB fixed point.
This behaviour has indeed been confirmed by direct numerical simulations of the CGL equation in the phase turbulence regime~\cite{Vercesi2024}, as shown in Fig.~\ref{fig:CGLE}.
The level lines of the correlation function $C(t,k)$ show a clear $z=1$ scaling.

Fig.~\ref{fig:CGLE}(b) displays the half decay time $\tau$ for different amplitude $\sigma$ of a microscopic Gaussian and delta-correlated noise $\eta$ added to the deterministic equation~\eqref{eq:CGLE_c1c2}. At small momenta $k$, one observes a crossover from the EW scaling (which is expected for not large enough system sizes) to the KPZ scaling  as the noise is increased (yielding a larger effective non-linearity). At even higher noise, the system enters a defect dominated phase,
which leads to a saturation of the decay time. However, remarkably, the large-momentum regime exhibits the IB scaling $z=1$, and this in a very robust manner, independently of the small-momentum behaviour. This shows that the IB universality intrinsically emerges in the CLG and KS equations, and thus appears in broader contexts than the KPZ equation. This opens exciting perspectives, to further explore the consequences of this finding within the CGL or KS equations.

A salient and related question is whether the IB regime can be observed in actual experiments. For experiments of interface growth, reasonably modelled by the KPZ equation, such as the beautiful liquid crystal platform~\cite{Takeuchi10,Takeuchi12}, the parameters --  surface tension $\nu$ or  non-linearity $\lambda$ -- are probably not easily tunable. One should try to estimate whether $g_\Lambda$ is large enough for the inviscid scaling to emerge at large momentum, and ensure that these large momenta can  be resolved. A perhaps more appropriate setting is provided by systems modelled by the CGL or KS equation. Among them is the exciton-polariton experiment previously mentioned, which realises driven-dissipative Bose-Einstein condensates~\cite{Fontaine2022Nat} and  stands as a  promising candidate. Indeed, the KPZ universal behaviour has recently been observed in such one-dimensional exciton-polariton condensate. Although the KPZ regime was found to extend  over a limited space-time window in the current setup, the parameters in this system can be varied to a certain extent, and it is possible that the range where the IB scaling should be observed  becomes accessible. Another intriguing possibility is  strongly interacting ultra-cold gases, described by the Bose-Hubbard model mentioned in the introduction, and studied through numerical simulations in Ref.~\cite{Fujimoto2020}. The findings of this work suggest that a IB regime could develop near half-filling, even though  a further theoretical support would be necessary to confirm that this system indeed belongs to the KPZ universality class, and that lowering the filling corresponds to increasing the effective non-linearity.

\section{Conclusion}

In this paper, we have given an overview of the non-perturbative aspects of the KPZ equation, which are the strong-coupling nature of the KPZ fixed point, and the existence of an infinite-coupling fixed point, the IB fixed point, in all dimensions, including $d=1$. These aspects can be accessed and characterised using the FRG approach. We have shown that even the simplest  approximation of the FRG already provides the complete phase diagram and associated scaling regimes of the KPZ equation. Then, we have demonstrated how, using more advanced approximations, very accurate results can be obtained, in particular, for the scaling functions associated with the correlation and response functions of the theory in the different scaling regimes. A salient point is that in the large-momentum limit, an exact form of the IB scaling function can be derived, relying only on taking this limit and exploiting  the extended symmetries of the KPZ equation.
As outlook, it would be interesting to investigate the KS and CLG equations using FRG to theoretically demonstrates the emergence of the IB fixed point during the flow from the microscopic scale to the KPZ fixed point, and shows that it indeed controls the behaviour of the correlation function at intermediate momenta. It would also be thrilling to reveal the signatures of the IB fixed point in actual experiments.

\section*{Acknowledgement}

LC wishes to thank all her collaborators, who have been involved in the work on the FRG analysis of the KPZ equation,
  H. Chat\'e, B. Delamotte,  N. Wschebor, T. Kloss, S. Mathey, D. Squizzato,
 M. Tarpin, C. Fontaine, F. Vercesi, L. Gosteva (in more or less chronological order).

%\bibliographystyle{SciPost_bibstyle}
%\bibliography{../../Biblio/bibNPRG.bib,../../Biblio/bibKPZ.bib,../../Biblio/bibBurgers.bib,../../Biblio/bibNS.bib, ../../Biblio/bibCGL-KS.bib}

\nolinenumbers

\end{document}